\documentclass[10pt,twocolumn,letterpaper]{article}

\usepackage[pagenumbers]{wacv} 

\usepackage{lettrine}
\usepackage[ruled,vlined]{algorithm2e}
\usepackage{multirow}

\usepackage{booktabs,tabularx,makecell}
\usepackage{pifont} 
\usepackage{array,tabularx,booktabs,rotating}
\newcolumntype{Y}{>{\centering\arraybackslash}X} 
\usepackage{colortbl}
\usepackage{makecell}

\definecolor{wacvblue}{rgb}{0.21,0.49,0.74}
\usepackage[pagebackref,breaklinks,colorlinks,allcolors=wacvblue]{hyperref}

\def\wacvPaperID{3002} 
\def\confName{WACV}
\def\confYear{2026}

\title{Geodesic Diffusion Models for Efficient Medical Image Enhancement}

\author{
Teng Zhang$^{1}$ \quad Hongxu Jiang$^{1}$ \quad Kuang Gong$^{2, 1}$ \quad
Wei Shao$^{3,1,4}$\footnotemark[1] \\
$^{1}$Department of Electrical and Computer Engineering, University of Florida, Gainesville, FL, USA \\
$^{2}$Department of Biomedical Engineering, University of Florida, Gainesville, FL, USA \\
$^{3}$Department of Medicine, University of Florida, Gainesville, FL, USA \\
$^{4}$Intelligent Clinical Care Center, University of Florida, Gainesville, FL, USA \\
{\tt\small \{zhangt,hongxu.jiang\}@ufl.edu \quad KGong@bme.ufl.edu \quad weishao@ufl.edu}
}

\begin{document}

\makeatletter
\twocolumn[{%
\begin{@twocolumnfalse}
\maketitle

\begin{center}
\includegraphics[width=16.5cm]{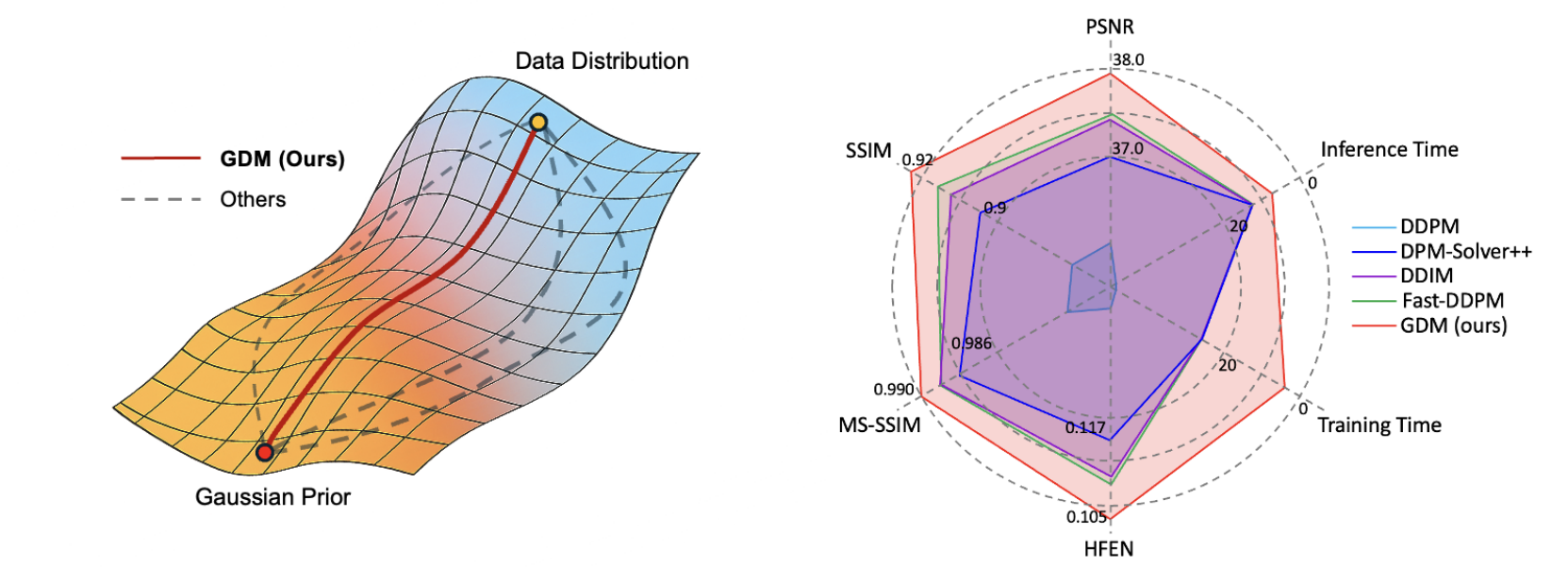}
\captionof{figure}{
Overview of Geodesic Diffusion Models (GDM). By constraining the diffusion process to follow geodesic paths with minimal energy in probability space, GDM improves both efficiency and fidelity. The radar plot shows performance on the CT denoising task, where GDM achieves higher accuracy and substantially faster training and sampling compared to prior diffusion-based methods.}
\label{fig:teaser}
\end{center}
\end{@twocolumnfalse}}
]

\begingroup
\renewcommand\thefootnote{\fnsymbol{footnote}}
\footnotetext[1]{Corresponding author: Wei Shao (\href{mailto:weishao@ufl.edu}{weishao@ufl.edu}).}
\footnotetext[2]{This work was supported by the Department of Medicine and the Intelligent Clinical Care Center at the University of Florida College of Medicine. The authors express their sincere gratitude to the NVIDIA AI Technology Center at the University of Florida for their invaluable feedback, technical guidance, and support throughout this project.}
\endgroup

\begin{abstract}
Diffusion models generate data by learning to reverse a forward process, where samples are progressively perturbed with Gaussian noise according to a predefined noise schedule. From a geometric perspective, each noise schedule corresponds to a unique trajectory in probability space from the data distribution to a Gaussian prior. However, prior diffusion models rely on empirically chosen schedules that may not be optimal. This inefficiency necessitates many intermediate time steps, resulting in high computational costs during both training and sampling. To address this, we derive a family of geodesic noise schedules corresponding to the shortest paths in probability space under the Fisher–Rao metric. Based on these schedules, we propose Geodesic Diffusion Models (GDMs), which significantly improve training and sampling efficiency by minimizing the energy required to transform between probability distributions. This efficiency further enables sampling to start from an intermediate distribution in conditional image generation, achieving state-of-the-art results with as few as 6 steps. We evaluated GDM on two medical image enhancement tasks: CT image denoising and MRI image super-resolution. Experimental results show that GDM achieved state-of-the-art performance while reducing training time by 20–30× compared to Denoising Diffusion Probabilistic Models (DDPMs) and 4–6× compared to Fast-DDPM, and accelerating sampling by 160–170× and 1.6×, respectively. These gains support the use of GDM for efficient model development and real-time clinical applications. Our code is publicly available at: \href{https://github.com/mirthAI/GDM-VE}{https://github.com/mirthAI/GDM-VE}.
\end{abstract}

\section{Introduction}
\label{sec:introduction}
Diffusion models \cite{ho2020denoising,song2020denoising,song2020score} have  emerged as powerful tools for high-quality image generation, surpassing generative adversarial networks in both fidelity and diversity \cite{dhariwal2021diffusion}. In the forward process, an image is progressively perturbed into Gaussian noise, while the reverse process is learned by a neural network that iteratively denoises random noise back into a high-quality image. Owing to their training stability and ability to generate high-fidelity images, diffusion models have been increasingly adopted in medical imaging \cite{pinaya2022brain,hu2023conditional,ozbey2023unsupervised,yu2025robust}. However, achieving state-of-the-art performance requires hundreds to thousands of denoising steps, which impose substantial computational costs in both training and sampling and thereby limit applicability in real-time medical settings.

We argue that this inefficiency is attributable to the suboptimal design of noise schedules. From a geometric standpoint, the forward diffusion process defines a trajectory on the manifold of Gaussian distributions, between a Gaussian concentrated at the data sample with near-zero variance and an isotropic Gaussian prior. The most efficient trajectory should correspond to a geodesic, that is, the shortest path between probability distributions under a suitable Riemannian metric. The Fisher–Rao metric is a natural choice, as it endows the statistical manifold with an intrinsic geometry by quantifying divergences induced by infinitesimal perturbations. However, most existing diffusion models \cite{ho2020denoising,nichol2021improved} rely on empirically specified schedules that deviate from geodesic paths. As a result, the diffusion trajectory passes through redundant intermediate states, which hinders information propagation and increases the computational cost of training and sampling. Defining noise schedules that align the diffusion process with geodesic trajectories under the Fisher–Rao geometry provides a principled approach to eliminating these inefficiencies while preserving image quality.

We introduce the Geodesic Diffusion Model (GDM), which constrains the diffusion process to follow a geodesic path between the data distribution and a Gaussian prior under the Fisher–Rao metric. This ensures that diffusion proceeds along the shortest trajectory in both the forward and reverse processes. With a geodesic noise schedule, GDM transforms the Gaussian prior into the target distribution along a engery-minimizing path without redundant intermediate states, thereby accelerating convergence during training. Since the diffusion path follows a geodesic in the Fisher–Rao manifold, fewer timesteps are required to solve the underlying differential equation, leading to significant improvement in sampling efficiency.

In this study, we focus on conditional medical image enhancement, where efficiency is particularly critical as clinical applications require substantially faster training and sampling. In this context, the conditional image provides structural information close to the target distribution, enabling sampling to begin from an intermediate noised state rather than pure noise. This truncated sampling strategy further reduces the number of required denoising steps. We evaluate GDM on two conditional medical image-to-image generation tasks: CT denoising using a single conditional image and MRI super-resolution using two conditional images. The key contributions of this work are summarized as follows:
\begin{enumerate}
\item \textbf{Geodesic Diffusion Model (GDM):} We propose a principled formulation of diffusion processes in which noise schedules follow geodesic trajectories under the Fisher–Rao metric, improving information propagation and generative fidelity.
\item \textbf{Computational efficiency:} GDM reduces training time by 20–30× compared to DDPM and 4–6× compared to Fast-DDPM, while accelerating sampling by 160–170× relative to DDPM and 1.6× relative to Fast-DDPM.
\item \textbf{Clinical relevance through conditional enhancement:} We adapt GDM to conditional medical image enhancement with truncated sampling, enabling state-of-the-art results in CT denoising and MRI super-resolution with substantially fewer timesteps.
\end{enumerate}

\section{Related Work}
\subsection{Fast Sampling of Diffusion Models}
To address the computational cost of diffusion models, most efforts have focused on accelerating sampling. Advanced solvers integrate the reverse SDE or ODE more accurately, reducing sampling to tens of steps without degrading image quality \cite{lu2022dpm,gonzalez2023seeds,zhao2023unipc,lu2025dpm}. Training-based approaches employ knowledge distillation to compress multi-step teachers into faster student models \cite{salimans2022progressive,luhman2021knowledge}. Rectified flow \cite{liu2022flow} straightens the generative trajectory through multiple rectifications, while Consistency Models \cite{song2023consistency} enforce trajectory invariance between timesteps, both accelerate sampling but incur higher training cost. Loss reweighting strategies have also been proposed to improve training efficiency \cite{hang2023efficient,hang2024improved}. Recently, Fast-DDPM \cite{jiang2025fast} was proposed for conditional image generation and improves both training and sampling efficiency by aligning the timesteps used in training and inference.

\begin{figure*}[hbt]
\centering
\includegraphics[width=0.85\textwidth]{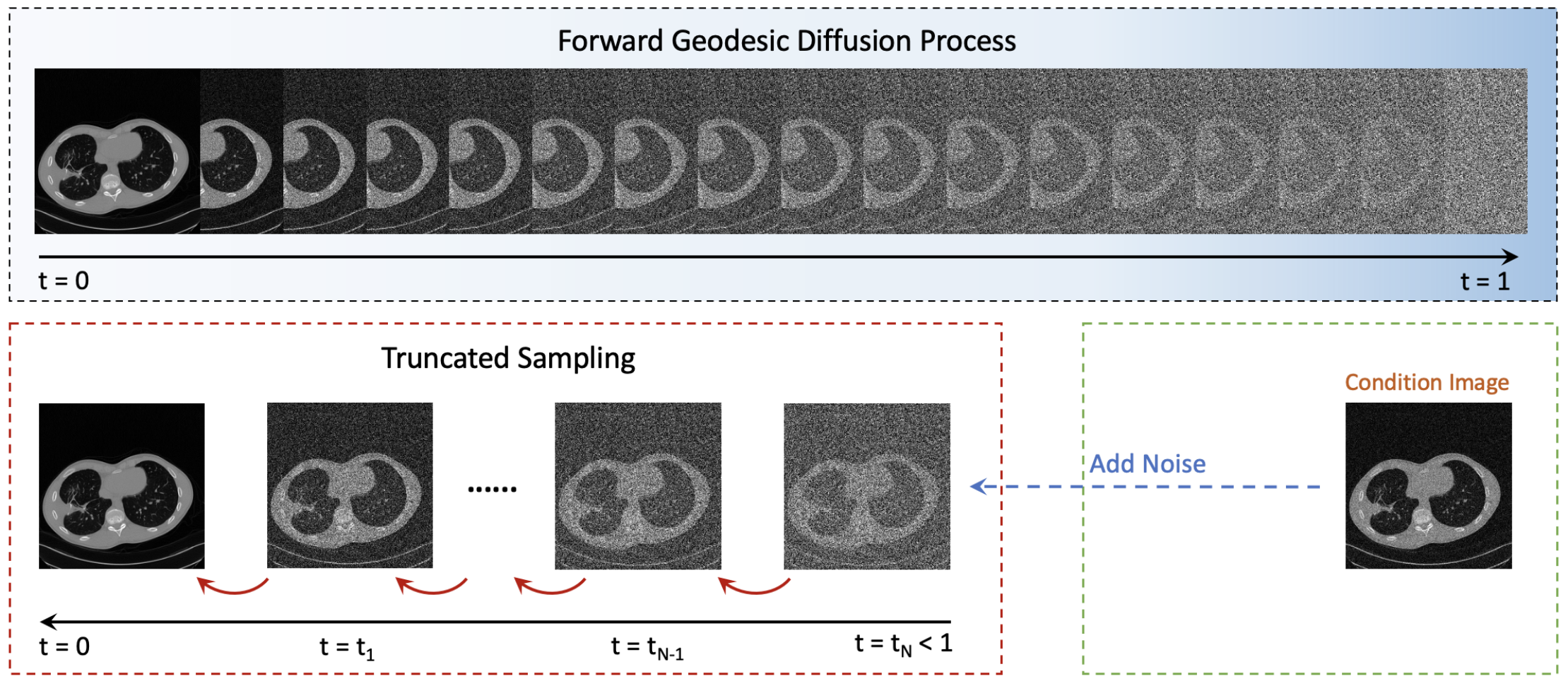}
\caption{Training and sampling pipeline of GDM. During training, the model learns an efficient continuous geodesic diffusion process. During sampling, noise is first added into condition image to simulate the intermediate image, and then only a few discrete steps are required to generate a high-quality image.}
\label{md}
\end{figure*}

\subsection{Noise Schedules for Diffusion Models}
The noise schedule specifies the amount of noise injected during diffusion, thereby determining the signal-to-noise ratio (SNR) at each timestep and strongly influencing both the efficiency and quality of the model. The design of the noise schedule is thus critical to the overall performance of diffusion models.  Fixed, hand-crafted schedules have been widely used in diffusion models, where the coefficients $(\alpha_t, \sigma_t)$ are predefined, and the SNR monotonically decreases across $t \in [0, 1]$. Common choices include Linear-$\beta$ \cite{ho2020denoising}, Cosine-$\alpha$ \cite{nichol2021improved}, and Sigmoid/Laplace schedules \cite{pmlr-v202-jabri23a, hang2024improved}. While straightforward to implement, such handcrafted schedules are not universally optimal. Prior work has shown that the optimal design of the noise schedule can be data- and task-dependent \cite{chen2023importance, hoogeboom2023simple}.

To overcome these limitations, adaptive or learnable schedulers have been proposed, parameterizing the noise schedule with trainable variables and optimizing jointly with the denoiser. For example, Variational Diffusion Models (VDM) \cite{kingma2021variational} employ a monotonic neural network $\gamma_{\eta}(t)$ to define the time-dependent variance and SNR. Diffusion Normalizing Flow (DiffFlow) \cite{zhang2021diffusion} learns the drift term in the forward SDE instead of fixing it, enabling adaptive noise schedules that better fit the data distribution.

\section{Methods}
\subsection{Mathematical Background}
\subsubsection{Diffusion Models}
The forward diffusion process for a high-quality image sample \( x_0 \sim p_{\text{data}} \) can be viewed as a continuous-time stochastic process that gradually transforms a near-zero–variance Gaussian distribution centered at \( x_0 \) into an isotropic Gaussian prior (pure noise). Formally, it is defined as
\begin{equation}
q(x_t \mid x_0) = \mathcal{N}\!\left(x_t; \alpha_t\, x_0, \sigma_t^2 \mathbb{I}_n\right), \quad t \in [0, 1],
\label{eq:forward}
\end{equation}
where \( \alpha_t \) and \( \sigma_t \) are continuous, scalar-valued functions of the normalized time index \( t \), \( \mathbb{I}_n \) denotes the \( n \times n \) identity matrix, and \( n \) is the data dimension. The signal-to-noise ratio,
\begin{equation}
\mathrm{SNR}(t) = \frac{\alpha_t^2}{\sigma_t^2},
\end{equation}
is typically a monotonically decreasing function, starting from a very large value at \( t = 0 \) (high signal, low noise) and approaching zero as \( t \to 1 \) (pure noise).

We can interpret the process in Equation~\eqref{eq:forward} as a trajectory in the space of Gaussian distributions with isotropic covariance, evolving from \( t = 0 \) to \( t = 1 \). This trajectory is fully determined by the pair \((\alpha_t, \sigma_t)\), commonly referred to as the \emph{noise schedule} of the diffusion process. A widely adopted choice is the variance-preserving (VP) schedule proposed in DDPM~\cite{ho2020denoising}. However, from a geometric perspective this schedule is suboptimal: under a Riemannian metric on the probability space, it fails to follow a geodesic (i.e., shortest path) between the endpoints, introducing inefficiencies in both training and sampling.



\subsubsection{The Fisher--Rao Metric}

The Fisher--Rao metric~\cite{costa2015fisher} is a natural Riemannian metric that measures path length in the space of parameterized probability distributions, enabling the identification of geodesics (shortest paths) between them. Given a probability distribution \( p(x; \theta_1, \dots, \theta_k) \) parameterized by \( \theta = (\theta_1, \dots, \theta_k)^T \), the Fisher information matrix \( I(\theta) \) is defined as
\begin{equation}
I(\theta) = \mathbb{E} \left[ \frac{\partial \log p(x; \theta)}{\partial \theta} 
\left( \frac{\partial \log p(x; \theta)}{\partial \theta} \right)^T \right].
\end{equation}
It quantifies the sensitivity of the distribution to parameter variations by capturing the expected curvature (second derivative) of the log-likelihood function. The Fisher information matrix induces a Riemannian metric on the statistical manifold, where the infinitesimal distance element is given by
\begin{equation}
ds = \sqrt{d\theta(t)^T I(\theta(t)) \, d\theta(t)}.
\end{equation}

\subsection{Geodesic Diffusion Models}

\subsubsection{Geodesic Noise Schedules}
Building on the Fisher--Rao metric, we derive a family of geodesic noise schedules that characterize the most efficient diffusion trajectories. 
Specifically, we consider the isotropic Gaussian distribution parameterized as $\theta(t) = (\alpha_t x_0, \sigma_t)^\top$ in Equation~\eqref{eq:forward}. 
Under this parameterization, the corresponding Fisher information matrix is
\begin{equation}
I(\theta) =
\begin{bmatrix}
\frac{1}{\sigma^2} I & 0 \\
0 & \frac{2n}{\sigma^2}
\end{bmatrix}.
\end{equation}

Using this metric, the squared distance of an infinitesimal change along the trajectory is
\begin{equation}
ds^{2} = \frac{A^{2}\dot{\alpha}^{2} + 2n\,\dot{\sigma}^{2}}{\sigma^{2}}\,dt^{2},
\label{eq:ds2}
\end{equation}
where $A:=\|x_0\|_2$ denotes the Euclidean norm of $x_0$. 
Accordingly, the kinetic energy of the path can be expressed as
\begin{equation}
L(\alpha,\sigma,\dot\alpha,\dot\sigma)
=\frac{1}{2}\Big(\frac{ds}{dt}\Big)^{2}
=\frac{A^2\dot\alpha^{\,2}+2n\dot\sigma^{\,2}}{2\sigma^2}.
\label{eq:FR_L}
\end{equation}

As shown in Appendix~1.1, a constant-speed geodesic is both length-minimizing and locally energy-minimizing. 
This implies that the trajectory must satisfy two conservation laws:

\textit{(i) Momentum along $\alpha$.} Since $L$ does not depend explicitly on $\alpha$,
\begin{equation}
p_\alpha:=\frac{\partial L}{\partial\dot\alpha}
=\frac{A^2}{\sigma^2}\dot\alpha=c \quad (\text{$c$ constant}),
\label{eq:momentum}
\end{equation}

\textit{(ii) Constant speed (energy).} Because $L$ is quadratic in velocities and independent of time,
\begin{equation}
\frac{A^2\dot\alpha^{\,2}+2n\dot\sigma^{\,2}}{\sigma^2}=k^2,
\qquad k>0.
\label{eq:energy}
\end{equation}

Solving Equations~\eqref{eq:momentum} and \eqref{eq:energy} (see Appendix~1.2) yields the closed-form geodesic scheduler:
\begin{equation}
\sigma(t)=Ar\,\mathrm{sech}(\theta_0-\delta t),
\label{eq:sigma}
\end{equation}
\begin{equation}
\alpha(t)=\alpha_0-r\sqrt{2n}\,[\tanh\theta_0-\tanh(\theta_0-\delta t)]\ .
\label{eq:alpha}
\end{equation}

Given boundary conditions $(\alpha_0,\sigma_0)$ at $t=0$ and $(\alpha_1,\sigma_1)$ at $t=1$, the parameters are determined as
\begin{equation}
\begin{split}
r=\frac{\sqrt{2n}}{2|\alpha_0-\alpha_1|}
\Biggl(&
\left(\frac{\sigma_1^2-\sigma_0^2}{A^2}
+\frac{(\alpha_0-\alpha_1)^2}{2n}\right)^{\!2} \\
&+\frac{2\sigma_1^2(\alpha_0-\alpha_1)^2}{A^2 n}
\Biggr)^{\!1/2},
\end{split}
\label{eq:r_closed}
\end{equation}
\begin{equation}
\theta_0=\mathrm{arcosh}\!\left(\tfrac{Ar}{\sigma_0}\right),
\label{eq:theta_0}
\end{equation}
\begin{equation}
\delta=\theta_0-\mathrm{arcosh}\!\left(\tfrac{Ar}{\sigma_1}\right).
\label{eq:delta}
\end{equation}

\begin{figure}
\includegraphics[width=\columnwidth]{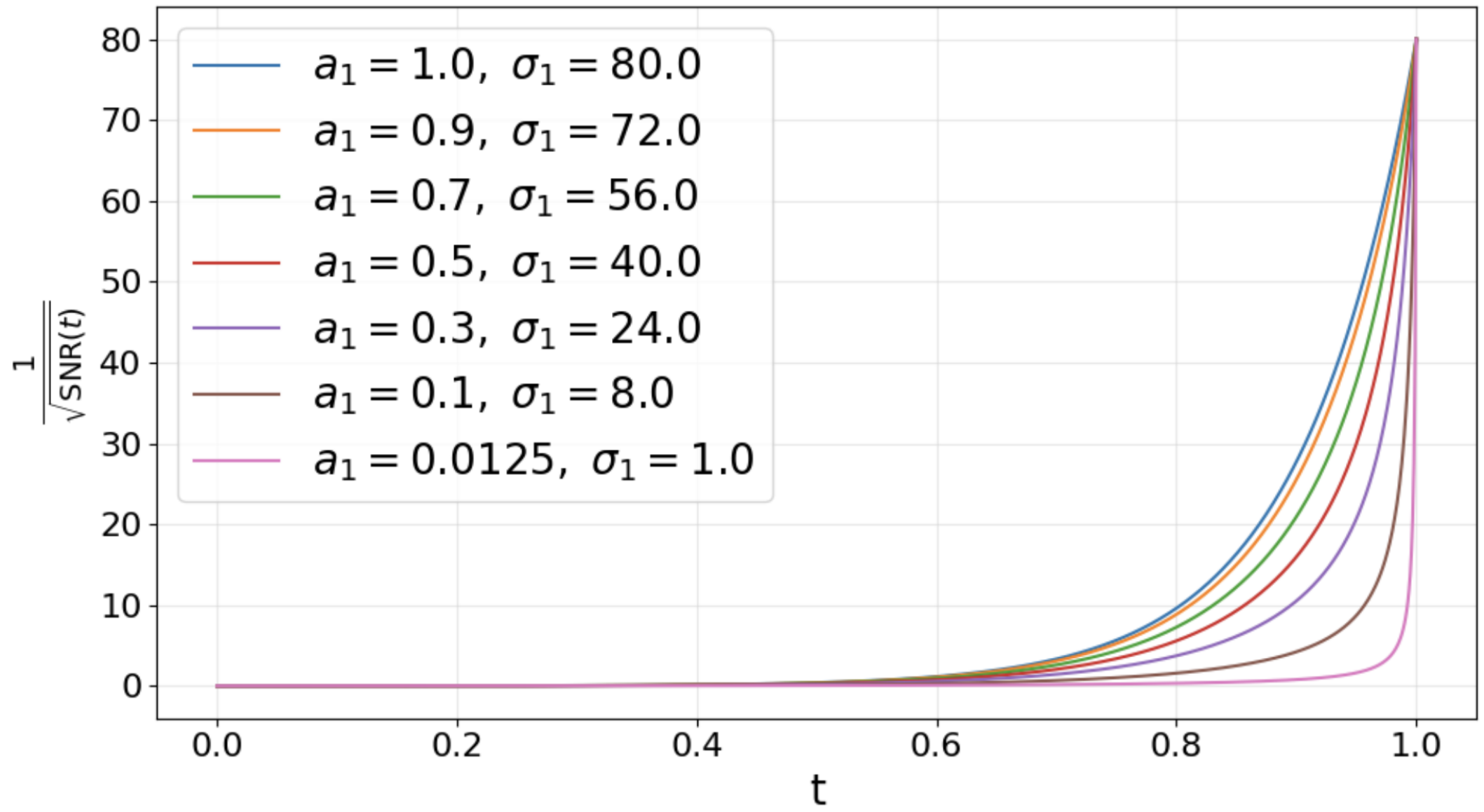}
\caption{A family of geodesic noise schedulers for different $\alpha_{1}$ and $\sigma_{1}$} 
\label{noise_schedulers_family}
\end{figure}

In practice, the forward diffusion process typically begins with a Gaussian distribution centered at $x_0$ and a very small standard deviation $\sigma_0$, with $\alpha_0 = 1$ and $\sigma_0$ commonly set to $0.002$~\cite{song2023consistency, karras2022elucidating}. 
As diffusion progresses, $\sigma_t$ increases while $\alpha_t$ decreases or remains unchanged, reducing the signal-to-noise ratio to a very small value (e.g., $1/6400$~\cite{song2023consistency, karras2022elucidating}). 
By selecting different $(\alpha_1, \sigma_1)$ pairs, one defines the Gaussian prior that serves as the terminal distribution of the forward process. 
Figure~\ref{noise_schedulers_family} illustrates a family of geodesic noise schedules corresponding to different priors, where each curve of $\frac{1}{\sqrt{\mathrm{SNR(t)}}} = \sigma_t / \alpha_t$
 represents the energy-minimizing path under the Fisher--Rao metric.

\subsubsection{Model Training}
Based on the geodesic noise schedules, we propose Geodesic Diffusion Models (GDMs) for conditional image enhancement, where the goal is to reconstruct a high-quality image \( x_0 \) from its conditional input \( c \), as illustrated in Figure~\ref{md}. We denote the joint distribution of images and conditions by \( p_{\text{joint}}(x_0, c) \), with the marginal distribution of conditions denoted as \( p_c(c) \).

During training, pairs \( (x_0, c) \sim p_{\text{joint}}(x_0, c) \) and timesteps \( t \sim U([0,1]) \) are randomly sampled. The clean image \( x_0 \) is then perturbed according to
\begin{equation}
x_t = \alpha_t x_0 + \sigma_t \epsilon,
\end{equation}
where \( \epsilon \sim \mathcal{N}(0, \mathbb{I}) \) is Gaussian noise, and \( \alpha_t, \sigma_t \) follow the geodesic schedule defined in Equations~\ref{eq:sigma}--\ref{eq:alpha}. A U-Net denoiser is trained by minimizing the following objective:
\begin{equation}
E_{(x_0, c), t, \epsilon} \left\| \epsilon - \epsilon_\theta \big(\alpha_t x_0 + \sigma_t \epsilon, c, t \big) \right\|^2,
\end{equation}
where \( \epsilon_\theta \) estimates the score function (up to a scaling factor) at each \( x_t \), i.e., \( \epsilon = - \sigma_t \nabla_{x_t} \log p(x_t) \). 

The complete training procedure for GDM is summarized in Algorithm~\ref{alg:training}.
\begin{algorithm}[th]
\SetAlgoLined
\caption{Training of Geodesic Diffusion Model}
\label{alg:training}
\Repeat{converged}{
  $t \sim U([0,1])$ \\
  $(x_0, c) \sim p_{\text{joint}}(x_0, c)$ \\
  $\epsilon \sim \mathcal{N}(0, \mathbb{I})$ \\
  Gradient descent: \\
  \hspace{1em} $\nabla_{\theta} \lVert \epsilon - \epsilon_{\theta}(\alpha(t)x_0 + \sigma(t)\epsilon, c, t) \rVert^2$
}
\end{algorithm}

\subsubsection{Model Sampling} 

In image enhancement tasks, the condition image $c$ is a degraded version of the high-quality target $x_0$ (e.g., a noisy image for denoising or a low-resolution image for super-resolution). Since $c$ already preserves structural and intensity information about $x_0$, the sampling process does not need to start from pure noise. This motivates truncated sampling, where generation begins at an intermediate noise level instead of the maximum. This motivates truncated sampling, where generation begins at an intermediate noise level instead of the maximum, thereby reducing computational cost while still preserving fidelity to the underlying data distribution.
We propose \emph{Geodesic Truncated Sampling} (GTS), which ensures that truncated sampling follows the constant-speed geodesic derived under the Fisher--Rao metric. By constraining the trajectory to remain on this geodesic, every intermediate image remains consistent with the diffusion process, leading to both efficiency and fidelity. Concretely, GTS starts by drawing random noise $\epsilon \sim \mathcal{N}(0, \mathbb{I})$ and sampling a condition $c$ (a single image for denoising or two images for super-resolution). The initialization is defined as
\[
x_{t_N} = \alpha_{t_N} c + \sigma_{t_N} \epsilon,
\]
where $x_{t_N} < 1$ and the ratio $\tfrac{\sigma_{t_N}}{\alpha_{t_N}}$ specifies the truncated noise level. Sampling is then initialized directly from this truncated state, rather than from pure noise, substantially reducing the number of required steps. The detailed procedure is summarized in Algorithm~\ref{alg:sampling}.



\begin{algorithm}[th]
	\SetAlgoLined
 \textbf{Hyperparameter:} number of sampling steps $ N $, start point of sampling $t_N$ \\
 $\epsilon \sim \mathcal{N}(0, \mathbb{I})$ \\
 $c \sim p_{\text{c}} (c)$ \\
 $x_{t_N} = \alpha_{t_N}c + \sigma_{t_N}\epsilon$ \\
\For{$k = N, \cdots, 1$  }{
$x_{t_{k-1}} = x_{t_k} + (t_{k-1} - t_k) \left[\frac{\dot{\alpha}_{t_k}}{\alpha_{t_k}} x_{t_k} + (\dot{\sigma}_{t_k} - \frac{\dot{\alpha}_{t_k}}{\alpha_{t_k}} \sigma_{t_k}) \epsilon_\theta (x_{t_k}, c, t_k)\right]$ \\
}
return $x_0$
 \caption{Sampling of Geodesic Diffusion Model}
\label{alg:sampling}
\end{algorithm}

\section{Experiments and Results}
We evaluated our proposed Geodesic Diffusion Model (GDM) on two medical image enhancement tasks: CT image denoising, conditioned on low-dose CT scans, and multi-image MRI super-resolution, conditioned on two neighboring MR slices.

\subsection{Experimental Details}
\subsubsection{Low-Dose and Normal-Dose Lung CT Dataset}
This publicly available dataset \cite{mccollough2021low} contains paired low-dose and normal-dose chest CT volumes. Normal-dose scans were acquired at standard clinical dose levels, while low-dose scans were simulated at 10\% of the routine dose. The dataset includes 48 patients, with 33 used for training, 5 for validation, and 10 for evaluation, resulting in 11,416, 1,795, and 3,501 pairs of 2D low-dose/normal-dose CT images, respectively. All images were resized to $256 \times 256$ pixels and normalized to the range $[-1, 1]$.

\subsubsection{Prostate MRI Dataset}
This publicly available dataset \cite{sonn2013targeted} consists of MRI volumes with an in-plane resolution of 0.547 mm × 0.547 mm and a through-plane resolution of 1.5 mm. For training, validation, and testing, we constructed triplets of three consecutive slices, using the first and third slices as input and the middle slice as the ground truth. In total, we extracted 6,689 triplets from 115 MRI volumes for training, 290 triplets from 5 volumes for validation, and 580 triplets from 10 volumes for evaluation. All slices were resized to $256 \times 256$ pixels and normalized to the range $[-1, 1]$.

\subsubsection{Evaluation Metrics}
To evaluate the quality of generated images, we used Peak Signal-to-Noise Ratio (PSNR), Structural Similarity Index Measure (SSIM), Multi-Scale SSIM (MS-SSIM), and High-Frequency Error Norm (HFEN). PSNR measures the ratio between maximum signal power and reconstruction error, with higher values indicating better fidelity. SSIM assesses perceived image quality by comparing luminance, contrast, and structural details, offering a more perceptually relevant evaluation than PSNR. MS-SSIM extends SSIM to multiple image scales, capturing both coarse and fine structures. HFEN emphasizes high-frequency components by applying a Laplacian-of-Gaussian filter to both generated and ground-truth images and measuring their difference.

\subsubsection{Implementation Details}
Our model adopts the U-Net denoiser architecture from DDIM \cite{song2020denoising}. We use the Adam optimizer with a learning rate of $2 \times 10^{-4}$ and a batch size of 16. All experiments were implemented in Python 3.10.6 with PyTorch 1.12.1 and run on four NVIDIA A100 GPUs with 80 GB memory each. Following prior work \cite{karras2022elucidating, song2023consistency}, we set $\alpha_0 = 1$, $\sigma_0 = 0.002$, $\alpha_1 = 1$, and $\sigma_1 = 80$. Truncated sampling was initialized at a noise level of $\sigma_{t_N} / \alpha_{t_N} = 3$.

\begin{table*}[hbt]
\caption{Quantitative results for CT image denoising and MRI super-resolution. 
The best results are shown in \textbf{\textcolor{red}{bold red}}, and the second-best in \textcolor{blue}{\underline{underlined blue}}. 
Training and inference times are reported as hours and seconds per volume, respectively. 
GDM converged after 70k iterations for CT denoising and 100k iterations for MRI super-resolution, compared with 2M and 400k iterations for DDPM and Fast-DDPM, respectively.}
\label{tab:results}
\centering
\small
\setlength{\tabcolsep}{3pt}
\begin{tabularx}{\textwidth}{c 
                                >{\centering\arraybackslash}m{5cm} 
                                >{\centering\arraybackslash}m{1.5cm} 
                                YYY
                                Y
                                Y
                                Y}
\toprule
\textbf{Task} & \textbf{Methods} & \textbf{\# Steps} & 
\textbf{PSNR}$\uparrow$ & \textbf{SSIM}$\uparrow$ & 
\textbf{\mbox{MS-SSIM}}$\uparrow$ & \textbf{HFEN}$\downarrow$ & \textbf{Training}$\downarrow$ & \textbf{Inference}$\downarrow$ \\
\midrule
\multirow{8}{*}{\rotatebox{90}{\textbf{Image Denoising}}}
 & RED-CNN~\cite{7947200}       & --    & 36.37 & 0.910 & 0.987 & 0.124 & \textbf{\textcolor{red}{3 h}}   & \textbf{\textcolor{red}{0.5 s}} \\
 & DU-GAN~\cite{9617604}       & --    & 36.25 & 0.902 & 0.986 & 0.128 & 20 h  & \textcolor{blue}{\underline{3.8 s}} \\
 & DDPM~\cite{ho2020denoising} & 1000  & 35.30 & 0.872 & 0.982 & 0.138 & 141 h & 1284 s \\
 & DDIM~\cite{song2020denoising} & 10   & 37.39 & 0.913 & \textcolor{blue}{\underline{0.989}} & 0.109 & 26 h  & 12.5 s \\
 & DPM-Solver++~\cite{lu2025dpm} & 10   & 37.02 & 0.904 & 0.988 & 0.114 & 26 h  & 12.5 s \\
 & Fast-DDPM~\cite{jiang2025fast} & 10  & 37.46 & \textcolor{blue}{\underline{0.916}} & \textcolor{blue}{\underline{0.989}} & \textcolor{blue}{\underline{0.108}} & 26 h  & 12.5 s \\
& \cellcolor{gray!15}\textbf{GDM w/o GTS (ours)}  & \cellcolor{gray!15}15 & \cellcolor{gray!15}\textcolor{blue}{\underline{37.85}} & \cellcolor{gray!15}\textbf{\textcolor{red}{0.923}} & \cellcolor{gray!15}\textbf{\textcolor{red}{0.990}} & \cellcolor{gray!15}\textbf{\textcolor{red}{0.103}} & \cellcolor{gray!15}\textcolor{blue}{\underline{4.25 h}} & \cellcolor{gray!15}18 s \\
& \cellcolor{gray!15}\textbf{GDM (ours)}      & \cellcolor{gray!15}6  & \cellcolor{gray!15}\textbf{\textcolor{red}{37.92}}   & \cellcolor{gray!15}\textbf{\textcolor{red}{0.923}}   & \cellcolor{gray!15}\textbf{\textcolor{red}{0.990}} & \cellcolor{gray!15}\textbf{\textcolor{red}{0.103}} & \cellcolor{gray!15}\textcolor{blue}{\underline{4.25 h}} & \cellcolor{gray!15}7.5 s \\
\midrule
\multirow{8}{*}{\rotatebox{90}{\textbf{Super Resolution}}}
 & miSRCNN~\cite{7115171}      & --    & 26.48 & 0.870 & 0.959 & 0.346 & \textbf{\textcolor{red}{1 h}}   & \textbf{\textcolor{red}{0.01 s}} \\
 & miSRGAN~\cite{SOOD2021101957} & --  & 26.79 & 0.880 & 0.965 & 0.331 & 40 h  & \textcolor{blue}{\underline{0.04 s}} \\
 & DDPM~\cite{ho2020denoising} & 1000  & 25.28 & 0.829 & 0.951 & 0.385 & 136 h & 222 s \\
 & DDIM~\cite{song2020denoising} & 10   & 26.58 & 0.877 & 0.963 & 0.340 & 26 h  & 2.3 s \\
 & DPM-Solver++~\cite{lu2025dpm} & 10   & 26.49 & 0.872 & 0.960 & 0.345 & 26 h  & 2.3 s \\
 & Fast-DDPM~\cite{jiang2025fast} & 10  & 27.05 & 0.887 & \textcolor{blue}{\underline{0.967}} & 0.314 & 26 h  & 2.3 s \\
& \cellcolor{gray!15}\textbf{GDM w/o GTS (ours)}  & \cellcolor{gray!15}15 & \cellcolor{gray!15}\textcolor{blue}{\underline{27.59}} & \cellcolor{gray!15}\textcolor{blue}{\underline{0.894}} & \cellcolor{gray!15}\textbf{\textcolor{red}{0.970}} & \cellcolor{gray!15}\textcolor{blue}{\underline{0.299}} & \cellcolor{gray!15}\textcolor{blue}{\underline{6.5 h}} & \cellcolor{gray!15}3.5 s \\
& \cellcolor{gray!15}\textbf{GDM (ours)}         & \cellcolor{gray!15}6  & \cellcolor{gray!15}\textbf{\textcolor{red}{27.61}} & \cellcolor{gray!15}\textbf{\textcolor{red}{0.896}} & \cellcolor{gray!15}\textbf{\textcolor{red}{0.970}} & \cellcolor{gray!15}\textbf{\textcolor{red}{0.297}} & \cellcolor{gray!15}\textcolor{blue}{\underline{6.5 h}} & \cellcolor{gray!15}1.4 s \\
\bottomrule
\end{tabularx}
\end{table*}

\subsection{CT Image Denoising}
To enable a comprehensive comparison, we evaluated six representative methods for CT denoising in terms of both image quality and computational efficiency: four diffusion-based models (DDPM \cite{ho2020denoising}, DDIM \cite{song2020denoising}, DPM-Solver++ \cite{lu2025dpm}, and Fast-DDPM \cite{jiang2025fast}), a convolution-based model (RED-CNN \cite{7947200}), and a GAN-based model (DU-GAN \cite{9617604}). As shown in Table~\ref{tab:results}, our GDM variants consistently achieve the best denoising performance across all metrics (PSNR, SSIM, MS-SSIM, and HFEN), outperforming CNN-, GAN-, and other diffusion-based baselines. In particular, GDM with GTS achieves the highest PSNR while maintaining top performance on SSIM, MS-SSIM, and HFEN.

Beyond accuracy, GDM also offers substantial efficiency improvements. Training time is reduced by $33\times$ compared to DDPM and $6\times$ compared to Fast-DDPM, requiring only 4.25 hours—comparable to training a convolutional model. At inference, GDM further accelerates runtime by $171\times$ relative to DDPM and $1.7\times$ relative to Fast-DDPM, achieving high-quality reconstructions with as few as six sampling steps. This dual advantage in accuracy and efficiency highlights GDM’s ability to overcome the efficiency bottleneck of diffusion models.

\begin{figure}[hbt]
\includegraphics[width=\columnwidth]{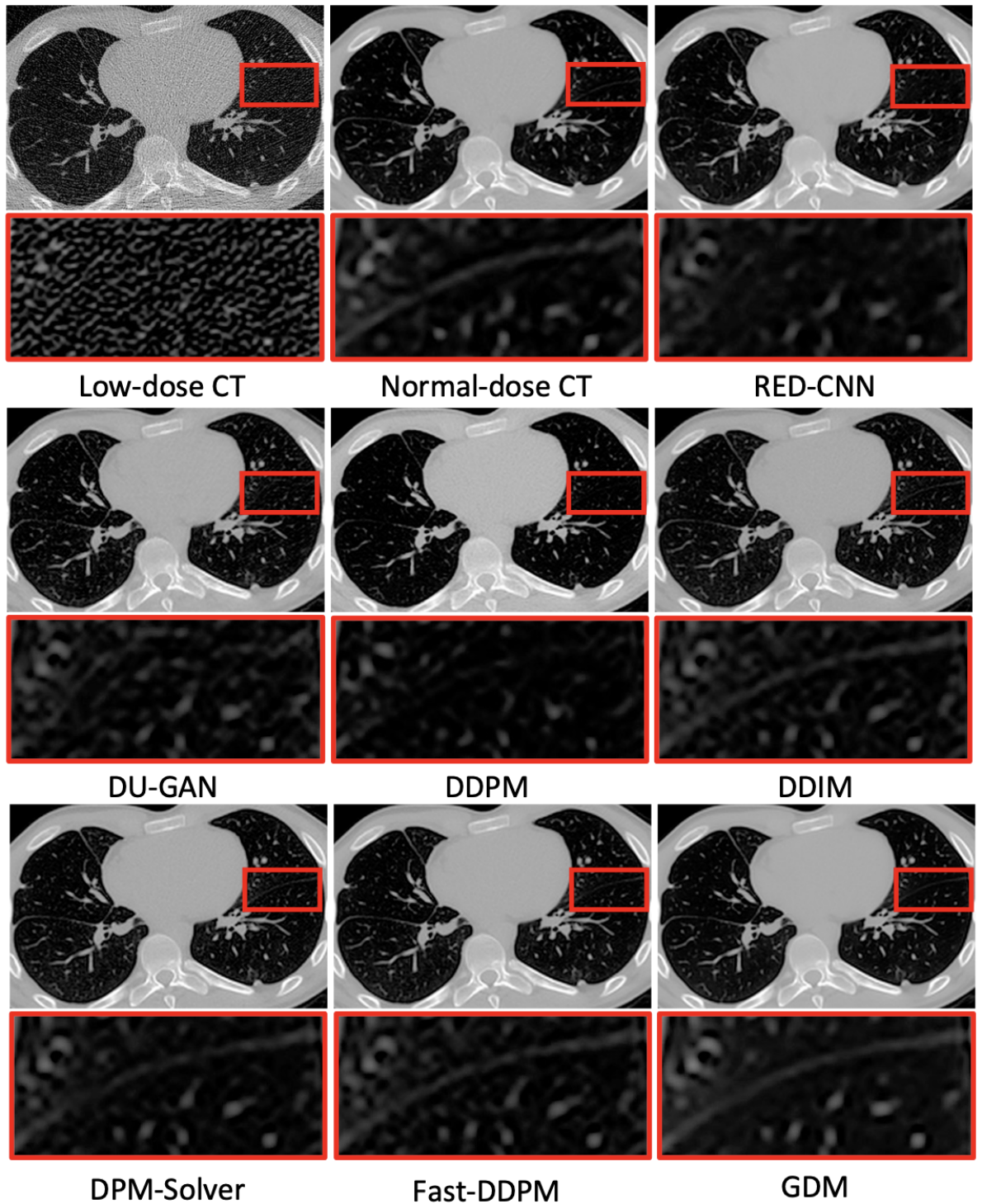}
\caption{Qualitative results for CT image denoising. Compared with CNN- (RED-CNN) and GAN-based (DU-GAN) methods, which blur or lose fine anatomical structures, diffusion-based models better preserve lung fissures. Our proposed GDM produces the sharpest and most continuous fissure reconstruction, closely matching the normal-dose reference.}
\label{Denoising}
\end{figure}

Figure~\ref{Denoising} shows qualitative comparisons on lung CT denoising. While RED-CNN and DU-GAN fail to preserve fine-grained anatomical structures such as lung fissures, diffusion-based models retain these details more effectively. Among them, GDM produces the sharpest and most continuous fissure reconstruction, closely resembling the normal-dose reference and supporting accurate downstream analysis.

\subsection{MR Image Super-Resolution}
For MRI super-resolution, we compared GDM against a convolution-based model (miSRCNN \cite{7115171}), a GAN-based model (miSRGAN \cite{SOOD2021101957}), and the same four diffusion-based baselines as in CT denoising. As shown in Table~\ref{tab:results}, GDM achieves state-of-the-art performance, surpassing all baselines. Even without truncated sampling, GDM ranks second overall, consistently outperforming prior methods. This demonstrates the effectiveness of geodesic trajectories in recovering subtle anatomical structures while suppressing noise. Beyond accuracy, GDM also delivers substantial efficiency gains, reducing training time by more than $20\times$ compared to DDPM and $4\times$ compared to Fast-DDPM, while pushing inference speed closer to that of lightweight regression-based models. This dual advantage underscores GDM’s potential as a practical solution for real-world MRI super-resolution, where both accuracy and scalability are critical.

\begin{figure*}[hbt]
\centering
\includegraphics[width=0.8\textwidth]{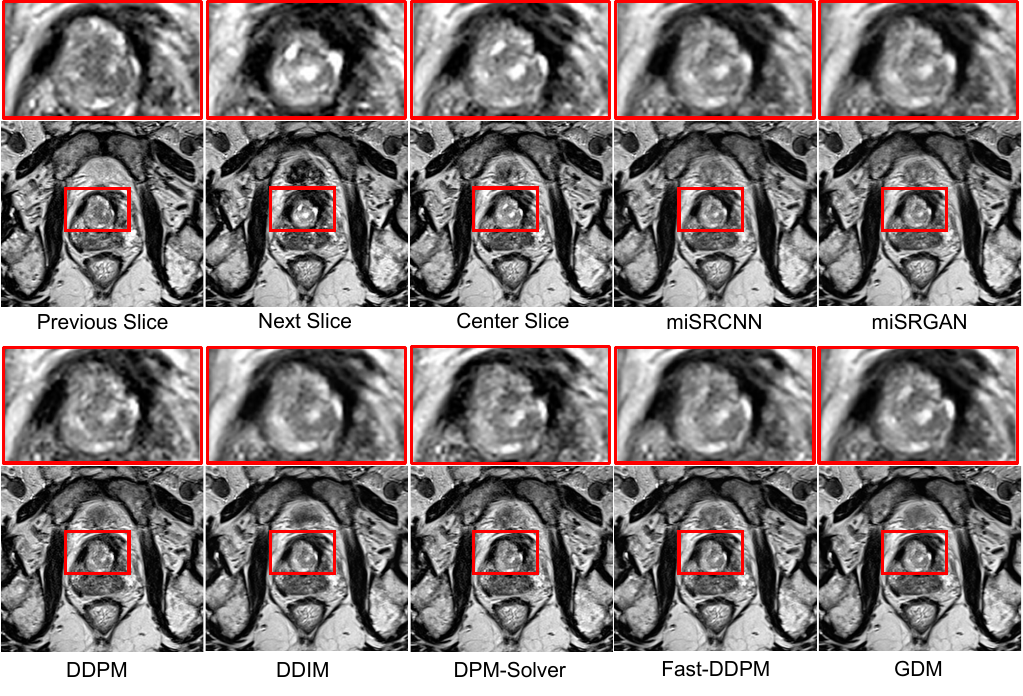}
\caption{Qualitative results of MRI image super-resolution. While CNN- (miSRCNN) and GAN-based (miSRGAN) methods produce blurry outputs and other diffusion models lose fine details, our proposed GDM reconstructs sharper anatomical structures and closely matches the ground-truth center slice.} \label{SR}
\end{figure*}

Figure~\ref{SR} provides qualitative comparisons on prostate MRI, where the center slice is reconstructed from adjacent slices. CNN-based models such as miSRCNN yield over-smoothed predictions, achieving higher PSNR but lower SSIM. Diffusion-based models preserve more detail, yet most fail to recover the bright regions in the upper-left corner. In contrast, GDM reconstructs these regions with sharp structural fidelity, closely matching the ground truth and demonstrating robustness in handling challenging anatomical variations.

\subsection{Ablation Study}
To evaluate the effectiveness and robustness of our proposed approach, we conducted three sets of ablation studies: varying the number of sampling steps, adjusting the truncated noise scale, and analyzing different noise schedules.

\paragraph{Impact of Sampling Steps.} Table~\ref{tab:step_analysis} examines how the number of sampling steps affects GDM performance under a fixed truncated noise level $\sigma_{t_N} / \alpha_{t_N}=3$. The results show that GDM maintains consistently high performance with 5--9 steps, achieving its best results at 6 steps. This demonstrates both the sampling efficiency of the geodesic trajectory and robustness to variation in step size. In contrast, reducing the number of steps below 5 causes a sharp collapse in PSNR and SSIM for both denoising and super-resolution, indicating that a minimal step budget is still required for reliable generation.

\begin{table}[hbt]
\caption{Effect of sampling steps on GDM performance with fixed truncated noise level $\sigma_{t_N} / \alpha_{t_N} = 3$. GDM maintains stable performance within 5--9 steps, peaking at 6, but collapses sharply when reduced below 5 steps.}
\label{tab:step_analysis}
\centering
\small
\setlength{\tabcolsep}{4pt}
\renewcommand{\arraystretch}{1.15}
\begin{tabularx}{\columnwidth}{c Y Y Y Y}
\toprule
\multirow{2}{*}{\textbf{Steps}} 
& \multicolumn{2}{c}{\textbf{Denoising}} 
& \multicolumn{2}{c}{\textbf{Super Resolution}} \\
\cmidrule(lr){2-3}\cmidrule(lr){4-5}
& PSNR$\uparrow$ & SSIM$\uparrow$ & PSNR$\uparrow$ & SSIM$\uparrow$ \\
\midrule
\rowcolor{gray!5} 9   & 37.51 & 0.915 & 27.59 & 0.894 \\
\rowcolor{gray!5} 8   & 37.61 & 0.917 & 27.57 & 0.894 \\
\rowcolor{gray!5} 7   & 37.68 & 0.920 & 27.54 & 0.894 \\
\rowcolor{gray!5} 6   & \textbf{37.92} & \textbf{0.923} & \textbf{27.61} & \textbf{0.896} \\
\rowcolor{gray!5} 5   & 37.53 & 0.917 & 26.33 & 0.877 \\
\rowcolor{red!5} 4    &  8.22 & 0.038 & 25.19 & 0.094 \\
\rowcolor{red!5} 3    &  5.50 & 0.011 &  5.64 & 0.031 \\
\bottomrule
\end{tabularx}
\end{table}

\paragraph{Impact of Truncated Noise Scale.}
Table~\ref{tab:noise_scale_analysis} examines the effect of different truncated noise levels $\sigma_{t_N}/\alpha_{t_N}$ on GDM performance. The results show that GDM maintains consistently high PSNR and SSIM across a wide range of noise scales, demonstrating strong robustness to this hyperparameter. Notably, setting $\sigma_{t_N}/\alpha_{t_N}=3$ provides the best balance, yielding peak performance while requiring fewer sampling steps. This robustness reduces the need for extensive hyperparameter tuning, making GDM more practical for real-world applications.

\begin{table}[t]
\caption{Effect of truncated noise levels $\sigma_{t_N}/\alpha_{t_N}$ on GDM performance. GDM remains stable across a wide range, with $\sigma_{t_N}/\alpha_{t_N}=3$ achieving the best trade-off between accuracy and efficiency.}
\label{tab:noise_scale_analysis}
\centering
\footnotesize
\rowcolors{3}{black!0}{black!0}
\begin{tabularx}{\columnwidth}{c YYYY}
\toprule
\multirow{2}{*}{$\sigma_{t_N}/\alpha_{t_N}$}
 & \multicolumn{2}{c}{\textbf{Denoising}}
 & \multicolumn{2}{c}{\textbf{Super Resolution}}\\
\cmidrule(lr){2-3}\cmidrule(lr){4-5}
 & PSNR$\uparrow$ & SSIM$\uparrow$ & PSNR$\uparrow$ & SSIM$\uparrow$ \\
\midrule
\rowcolor{gray!5} 80 & 37.85 & \textbf{0.923} & 27.61 & 0.894 \\
\rowcolor{gray!5} 60 & 37.89 & \textbf{0.923} & 27.59 & 0.895 \\
\rowcolor{gray!5} 50 & 37.89 & 0.922 & 27.61 & 0.894 \\
\rowcolor{gray!5} 20 & 37.91 & \textbf{0.923} & \textbf{27.64} & 0.895 \\
\rowcolor{gray!5} 10 & 37.91 & \textbf{0.923} & 27.63 & 0.895 \\
\rowcolor{gray!5} 5  & 37.88 & \textbf{0.923} & 27.57 & 0.895 \\
\rowcolor{gray!5} 3  & \textbf{37.92} & \textbf{0.923} & 27.61 & \textbf{0.896} \\
\rowcolor{gray!5} 2   & 37.73 & 0.918 & 27.50 & 0.894 \\
\rowcolor{gray!5} 1   & 37.50 & 0.916 & 27.55 & \textbf{0.896} \\
\bottomrule
\end{tabularx}
\end{table}

\paragraph{Impact of Noise Schedule.}
We evaluated the sensitivity of GDM to the geodesic noise scheduling trajectory by varying the terminal parameters $(\alpha_{1}, \sigma_{1})$. Although all configurations share the same initial and final signal-to-noise ratios (SNR), they define different geodesic trajectories starting from distinct Gaussian priors, as illustrated in Figure~\ref{noise_schedulers_family}. As shown in Table~\ref{tab:alpha_sigma_t1}, GDM remains remarkably stable across the first five configurations, consistently reaching around 37.9 dB PSNR for denoising and 27.6 dB for super-resolution with nearly identical SSIM values. This highlights strong robustness to schedule variations and validates the geometric principles underlying GDM.  

In contrast, performance collapses under the final two configurations, where the effective noise rises too abruptly near $t=1$. Such late-stage noise injection disrupts learning signals during the last diffusion steps, severely degrading reconstruction quality. These results provide practical guidance: noise schedules should ensure smooth growth while preserving energy-minimizing trajectories to maximize the effectiveness of diffusion models.

\begin{table}[hbt]
\caption{Effect of varying terminal parameters $(\alpha_{1}, \sigma_{1})$ at $t=1$ on GDM performance. GDM remains stable across moderate schedule variations, but quality degrades when noise increases too abruptly near the end of the diffusion process.}
\label{tab:alpha_sigma_t1}
\centering
\small
\setlength{\tabcolsep}{5pt}
\renewcommand{\arraystretch}{1.15}
\begin{tabularx}{\columnwidth}{c c Y Y Y Y}
\toprule
\multirow{2}{*}{$\sigma_{1}$} & \multirow{2}{*}{$\alpha_{1}$} 
& \multicolumn{2}{c}{\textbf{Denoising}} 
& \multicolumn{2}{c}{\textbf{Super Resolution}} \\
\cmidrule(lr){3-4}\cmidrule(lr){5-6}
& & PSNR$\uparrow$ & SSIM$\uparrow$ & PSNR$\uparrow$ & SSIM$\uparrow$ \\
\midrule
\rowcolor{gray!5} 80 & 1      & \textbf{37.92} & \textbf{0.923} & \textbf{27.61} & \textbf{0.896} \\
\rowcolor{gray!5} 72 & 0.9    & 37.91 & 0.922 & \textbf{27.61} &  \textbf{0.896} \\
\rowcolor{gray!5} 56 & 0.7    & 37.90 & 0.922 & 27.59 &  \textbf{0.896} \\
\rowcolor{gray!5} 40 & 0.5    & 37.90 & 0.922 & 27.57 &  \textbf{0.896} \\
\rowcolor{gray!5} 24 & 0.3    & 37.90 & \textbf{0.923} & 27.56 &  \textbf{0.896} \\
\rowcolor{red!5} 8  & 0.1    & 34.09 & 0.762 & 26.30 & 0.866 \\
\rowcolor{red!5} 1  & 0.0125 & 21.15 & 0.501 & 17.34 & 0.474 \\
\bottomrule
\end{tabularx}
\end{table}

\section{Discussion and Conclusion}
This paper introduced Geodesic Diffusion Models (GDMs), which optimize the diffusion process by constraining transformations between data distributions to follow the shortest path in probability space. This principled formulation reduces inefficiencies and enables a more direct, computationally effective generative process. Experimental results show that GDM achieves state-of-the-art performance across medical image enhancement tasks while substantially reducing computational overhead, underscoring its potential to advance medical imaging research and support real-time clinical applications.

Future work will extend GDM to 3D image generation and investigate its applicability to multi-modal medical imaging, enabling deeper integration across diverse data sources. The flexibility and efficiency of GDM establish it as a promising direction for future developments in diffusion-based generative modeling.

{
    \small
    \bibliographystyle{ieeenat_fullname}
    \bibliography{ref}
}

\clearpage                
\appendix                  
\setcounter{equation}{0}
\setcounter{figure}{0}
\setcounter{table}{0}
\renewcommand{\theequation}{S\arabic{equation}}
\renewcommand{\thefigure}{S\arabic{figure}}
\renewcommand{\thetable}{S\arabic{table}}

\section*{Supplementary Material for Geodesic Diffusion Models}

\section{Extended Mathematical Derivations}

\subsection{Energy--Length Relationship under the Fisher--Rao Metric}

The squared distance of an infinitesimal change is
\begin{equation}
ds^{2} \;=\; \frac{A^{2}\dot{\alpha}^{2} + 2n\,\dot{\sigma}^{2}}{\sigma^{2}}\,dt^{2}.
\label{eq:ds2}
\end{equation}
The corresponding kinetic energy (Lagrangian) is
\begin{equation}
L(\alpha,\sigma,\dot\alpha,\dot\sigma)
=\frac{1}{2}\Big(\frac{ds}{dt}\Big)^{2}
=\frac{A^2\dot\alpha^{\,2}+2n\dot\sigma^{\,2}}{2\sigma^2}.
\label{eq:FR_L}
\end{equation}
Define the \emph{speed} of the path as
\begin{equation}
v(t)\;:=\;\frac{ds}{dt}
=\sqrt{\frac{A^2\dot\alpha^{\,2}+2n\dot\sigma^{\,2}}{\sigma^2}} \;\ge 0.
\label{eq:speed}
\end{equation}
Then the path length and energy on $[0,1]$ are
\begin{equation}
\ell \;=\;\int_{0}^{1} v(t)\,dt,\qquad
E \;=\; \frac12\int_{0}^{1} v(t)^{2}\,dt.
\label{eq:length_energy_v}
\end{equation}

\paragraph{Cauchy--Schwarz inequality.}
Applying Cauchy--Schwarz on $[0,1]$ with $f=v$ and $g=1$ gives
\begin{equation}
\ell^2
= \Big(\!\int_0^1 v\,dt\!\Big)^2
\le \int_0^1 v^2\,dt
= 2E.
\end{equation}
Thus
\begin{equation}
E \;\ge\; \frac{\ell^{2}}{2}.
\label{eq:E_ge_l2_over_2}
\end{equation}

\paragraph{Equality condition.}
Equality holds iff $v(t)$ is constant. In this case $v(t)\equiv c\ge 0$, so
$\ell = \int_{0}^{1} v\,dt = c$, and
$E = \tfrac12\int_{0}^{1} c^2\,dt = \tfrac12\ell^2$,
matching \eqref{eq:E_ge_l2_over_2}.

\paragraph{General interval $[a,b]$.}
If the time interval has length $T=b-a$, the inequality generalizes to
\begin{equation}
E \;\ge\; \frac{\ell^{2}}{2T},
\end{equation}
with equality iff $v$ is constant.

\paragraph{Consequences.}
Fixing endpoints in a convex neighborhood:
\begin{enumerate}
\item For any given geometric curve of length $\ell$, energy is minimized by its constant–speed parameterization, with minimal value $E_{\min}=\ell^2/2$.
\item Among all nearby curves, the unique geodesic minimizes length $\ell$. Endowed with constant speed, its energy is $E=\ell_{\mathrm{geo}}^2/2$, and no competitor can achieve lower energy.
\end{enumerate}
Thus a constant–speed geodesic is simultaneously length– and energy–minimizing.

\subsection{Closed-Form Geodesic Noise Schedule}

We now derive a closed-form geodesic noise schedule from the Fisher--Rao Lagrangian.

A path is geodesic if it makes the energy functional
$E=\int_0^1 L\,dt$ stationary under endpoint-fixed perturbations. Choosing $t$ as an affine (constant–speed) parameter yields two conservation laws.

\textit{(i) Momentum along $\alpha$.}
Since $L$ does not depend on $\alpha$,
\begin{equation}
p_\alpha:=\frac{\partial L}{\partial\dot\alpha}
=\frac{A^2}{\sigma^2}\dot\alpha=c \quad (\text{$c$ constant}),
\label{eq:momentum}
\end{equation}
hence $\dot\alpha = c\,\sigma^2/A^2$.

\textit{(ii) Constant speed (energy).}
Because $L$ is quadratic in velocities and time-independent,
\begin{equation}
\frac{A^2\dot\alpha^{\,2}+2n\dot\sigma^{\,2}}{\sigma^2}=k^2,
\qquad k>0.
\label{eq:energy}
\end{equation}

Combining \eqref{eq:momentum} and \eqref{eq:energy} eliminates $\dot\alpha$, yielding the ODE
\begin{equation}
\dot\sigma=\alpha_*\,\sigma\sqrt{\,1-B^2\sigma^2\,},
\quad \alpha_*:=\frac{k}{\sqrt{2n}},\ \ 
B:=\frac{|c|}{Ak}.
\label{eq:sigma_ode}
\end{equation}
Separating variables and integrating gives
\begin{equation}
\sigma(t) = \lambda\,\mathrm{sech}(\theta_0-\delta t),
\quad
\lambda := \tfrac{1}{B}=\tfrac{Ak}{|c|},\;\;
\delta := \tfrac{k}{\sqrt{2n}}.
\label{eq:sigma_closed}
\end{equation}
From \eqref{eq:momentum}, integration yields
\begin{equation}
\alpha(t)=\alpha_0-\frac{c\lambda^2}{A^2\delta}\,
\big[\tanh\theta_0-\tanh(\theta_0-\delta t)\big].
\label{eq:alpha_closed}
\end{equation}

Defining $r:=k/|c|>0$ gives $\lambda=Ar$. Substituting, the geodesic scheduler is
\begin{equation}
\sigma(t)=Ar\,\mathrm{sech}(\theta_0-\delta t),
\label{eq:sigma}
\end{equation}
\begin{equation}
\alpha(t)=\alpha_0-r\sqrt{2n}\,[\tanh\theta_0-\tanh(\theta_0-\delta t)].
\label{eq:alpha}
\end{equation}
Given boundary conditions
$(\alpha_0,\sigma_0)$ at $t=0$ and $(\alpha_1,\sigma_1)$ at $t=1$, we have
\begin{equation}
\begin{split}
r=\frac{\sqrt{2n}}{2|\alpha_0-\alpha_1|}
\Biggl(&
\left(\frac{\sigma_1^2-\sigma_0^2}{A^2}
+\frac{(\alpha_0-\alpha_1)^2}{2n}\right)^{\!2} \\
&+\frac{2\sigma_1^2(\alpha_0-\alpha_1)^2}{A^2 n}
\Biggr)^{\!1/2},
\end{split}
\label{eq:r_closed}
\end{equation}
\begin{equation}
\theta_0=\mathrm{arcosh}\!\left(\tfrac{Ar}{\sigma_0}\right),
\label{eq:theta_0}
\end{equation}
\begin{equation}
\delta=\theta_0-\mathrm{arcosh}\!\left(\tfrac{Ar}{\sigma_1}\right).
\label{eq:delta}
\end{equation}

\paragraph{Special case.}
If $\alpha_t$ remains fixed ($\alpha_0=\alpha_1$), then $\dot\alpha\equiv0$ and $c=0$. From \eqref{eq:energy},
$\dot\sigma=\delta\,\sigma$ with $\delta=\tfrac{k}{\sqrt{2n}}$, giving the exponential scheduler:
\begin{equation}
\sigma(t)=\sigma_0\,e^{\delta t}
=\sigma_0\left(\frac{\sigma_1}{\sigma_0}\right)^{t},
\label{eq:exp}
\end{equation}
a degenerate case of the general geodesic solution.


\end{document}